\documentclass[singlespacing]{elsart}

\usepackage{amssymb}
\usepackage{graphicx}

\begin{document}

\begin{frontmatter}

\newtheorem{Lemma}{Lemma}
\newtheorem{Theorem}{Theorem}
\newtheorem{Proposition}{Proposition}
\newtheorem{Corollary}{Corollary}
\newtheorem{Definition}{\hspace{1.5cm}}[section]

\title{A cellular automata model for avascular solid tumor growth under the effect of therapy}

\author  {E. A. Reis}
\ead{eareisphysics@gmail.com}
\author  {L. B. L. Santos}
\ead{santosl@ufba.br}
\author  {S. T. R. Pinho}
\ead{suani@ufba.br}
\address  {Instituto de F\'{\i}sica,
Universidade Federal da Bahia, 40210-340, Salvador, Brazil}

\date{\today}

\begin{abstract}

Tumor growth has long been a target of investigation within the
context of mathematical and computer modelling. The objective of
this study is to propose and analyze a two-dimensional probabilistic
cellular automata model to describe avascular solid tumor growth,
taking into account both the competition between cancer cells and
normal cells for nutrients and/or space and a time-dependent
proliferation of cancer cells. Gompertzian growth, characteristic of
some tumors, is described and some of the features of the
time-spatial pattern of solid tumors, such as compact morphology
with irregular borders, are captured.  The parameter space is
studied in order to analyze the occurrence of necrosis and the
response to therapy. Our findings suggest that transitions exist
between necrotic and non-necrotic phases (no-therapy cases), and
between the states of cure and non-cure (therapy cases). To analyze
cure, the control and order parameters are, respectively, the
highest probability of cancer cell proliferation and the probability
of the therapeutic effect on cancer cells. With respect to patterns,
it is possible to observe the inner necrotic core and the effect of
the therapy destroying the tumor from its outer borders inwards.

\end{abstract}


\end{frontmatter}

\smallskip\

\maketitle

\section{INTRODUCTION}

Neoplastic diseases are the cause of 7 million deaths annually or 12%
of deaths worldwide \cite{Who}. Mathematical and computer modelling
may lead to greater understanding of the dynamics of cancer
progression in the patient \cite{Preszosi} \cite{MMC}. These
techniques may also be useful in selecting better therapeutic
strategies by subjecting available options to computer testing ({\it
in silico}). Continuous models have been proposed to describe the
stages of tumor growth since the middle of the 20th century
\cite{Wheldon}. Initially, a tumor grows exponentially (linear
rate). After this transient stage, the growth rate decreases and a
steady state is attained, due to several factors including a lack of
nutrients and hypoxia. This nonlinear behavior characterizes
avascular tumor growth when neovascularization has not yet been
triggered. The decelerating avascular growth may be guided by
different rules such as, for example, Gompertzian and logistic
functions.  Gompertzian growth has been one of the most studied
decelerating tumor growth over the past 60 years \cite{Laird},
\cite{Demicheli} \cite
 {Brunton}. It is found, for example, in some solid tumors such as
breast carcinomas \cite{Clare} \cite{Spratt}. It is also observed in
tumors {\it in vitro} \cite{Bellomo}
  \cite{Guiot}.

Although continuous models are capable of describing the behavior of
tumor growth, it would appear more reasonable to adopt a discrete
approach when describing the prevascular stage. Due to the fact that
the angiogenic process has not triggered early tumor growth, few
cancer cells are present and growth depends predominantly on the
interactions of these cells with adjacent cells and with the
environment \cite{Patel}. In addition, in the discrete approach, it
is easier to capture the time-spatial pattern generated by the model
in order to compare it with actual patterns \cite{Castro}. Some
cellular automata models \cite{Galam} \cite{Shen} and hybrid
cellular automata \cite{Patel} \cite{Dormann} \cite{Gerlee} have
been proposed to study tumor growth.

Another important topic that is analyzed in mathematical models is
the response to therapy, including how tumor growth changes under
the effect of a drug \cite{Rygaard}. The focus is directed towards
identifying the optimal therapy to maximize the effect on cancer
cells and minimize the effect on normal cells \cite{Martin}
\cite{Swan}. Although various continuous chemotherapy models exist
\cite{Murray}, \cite{Costa}, \cite{Matveev}, to the best of our
knowledge the majority of the discrete models cited in the
literature have not yet been used to investigate this topic.

The objective of this work is to propose a two-dimensional cellular
automata model consisting of 4 states (empty site, normal cell,
cancer cell or necrotic tumor cell) to describe avascular tumor
growth. It should be emphasized that in this study the term tumor
growth is used to refer to the number of cancer cells rather than
the volume of the tumor; in other words, it is assumed that the
tumor volume is proportional to the number of cancer cells
\cite{Wheldon}. Assuming that the angiogenic process has not yet
been triggered, there is no increase in nutrients, which are
uniformly distributed over the lattice. In this simple model, some
relevant processes involved in the prevascular phase of tumor growth
are assessed: a dynamic proliferation of cancer cells and the
competition between normal cells and cancer cells for nutrients
and/or space. Since necrosis is often present in the prevascular
stage of tumor growth \cite{Bellomo} \cite{Adam}, the possibility of
necrosis in the model must also be taken into consideration.
Finally, the effect of therapy is included in order to investigate
whether the system evolves to a state of cure.

This paper is organized as follows. In section \ref{sec2}, the model
is presented, together with its local rules, parameters and the
scope of the algorithm. Section \ref{sec3} describes the simulated
time series of cell density in the presence or absence of treatment,
and shows the features of the time-spatial pattern of simulated
solid tumors. In section \ref{sec4}, the parameter related to the
process of necrosis and the effect of therapy is analyzed. Finally,
in section \ref{sec5}, our results are discussed from the point of
view of the phenomenon and some concluding remarks are made.

\section{THE MODEL}
\label{sec2}

We propose a two-dimensional ($L \times L$) cellular automata model
\cite{Wolfram} under periodic boundary conditions, using a Moore
neighborhood with a radius of 1.  At the initial condition ($t=0$),
there is only one cancer cell (to ensure better visualization, this
was taken from the center of the lattice). Since the intention it to
model a non-viral tumor, normal cells would not be transformed into
cancer cells with the exception of the cancer cell that triggers
tumor growth at $t = 0$ \cite{Thecell}. The lattice represents a
tissue sample; there is a cell in each site that may be in one of
four states: normal cell (NoC), cancer cell (CC), necrotic cell
(NeC) or empty site (ES). We assume that the nutrients are uniformly
available over the lattice. In this respect, lack of space is
identified with lack of nutrients in our model.

There is a growth potential $P_c$ value associated with each normal
or cancer cell. Although the two-dimensional character of the model
mimics the {\it in vitro} situation, the growth potential of cancer
cells simulates the three-dimensional tumor {\it in vivo} in the
sense that it represents the total number of cancer cells, i.e., in
addition to the cancer cells on the lattice, the cancer cells
generated by these cells.

The local rules are such that:

\begin{itemize}

 \item [] (i) The initial value of the mitotic probability of cancer cells is
 represented by a parameter $p_0$ that measures the available resources at
 the beginning of the tumor. After that, it decreases by a factor $\Delta
 p_{mitot}$ until reaching the null value:
\begin{equation}
\label{pmitot}
 \Delta p_{mitot} = \exp \left[-
\left(\frac{n_{noc}(t)}{n_{cc}(t)}\right)^2 \right ]
\end{equation}

\noindent  in which $n_{noc}$ and $n_{cc}$ are the number of normal
and cancer cells at time $t$, respectively. As shown, $\Delta
p_{mitot}(t)$ depends only on the dynamics, setting up a feedback
inhibition mechanism \cite{Castro}: as the tumor grows, $\Delta
p_{mitot}(t)$ decreases because of the combined effect of the
decrease in the number of normal cells and the increase in the
number of cancer cells. In order to intensify the effect of this
mechanism (see \cite{Lobato}), an exponent 2 in equation
(\ref{pmitot}) is considered. Since there is no new available source
of nutrients and/or space, it decreases as the density of cancer
cells increases because the available nutrients and/or space are
reduced. The effect of the proliferation of cancer cells is that
their growth potential increases by a unit at each time step.

\vspace{0.3cm}

\item [] (ii) The cancer cells compete with normal cells for the empty
sites, depending on the potential growth of neighboring cells.
According to the majority rule, a normal cell is displaced by a
cancer cell following local battles occurring between healthy and
cancerous cells [15].

\vspace{0.3cm}

\item [] (iii) If the growth potential of a cancer cell reaches a
threshold value that is a fraction $f$  of the lattice size $L$, it
becomes necrotic and its growth potential falls to zero.

\vspace{0.3cm}

\item [] (iv) Both normal and cancer cells may die, with probabilities
$p_{drugn}$ and $p_{drugc}$, respectively, due to the continuous
infusion of a drug that is applied after tap time steps; in this
case, the site becomes empty.

\vspace{0.3cm}

\item [] (v) if there are no cancer cells in the neighborhood of a dead
cell (empty site), regeneration of normal cells occurs; if the
cancer cells in the neighborhood of a necrotic cell die as a result
of the therapy, the necrotic cell is eliminated.

\end{itemize}

The algorithm was computationally implemented in FORTRAN 77 in
accordance with the following steps: input data; calculate $\Delta
p_{mitot}(t)$; identify the state of the cell (choose one of the
subroutines: normal cell (NoC), cancer cell(CC), necrotic cell(NeC),
empty site (ES); update the cells of the lattice; after N
iterations, output data.

 The input data are the
following cellular automata (CA) parameters:

\begin{itemize}

\item [] 1) The spatial parameters: lattice size L; necrosis threshold
fraction $f$ of lattice size;

\vspace{0.3cm}

\item [] 2) The temporal parameters: the length of the time series t
final and the initial time of therapy infusion $t_{ap}$;

\vspace{0.3cm}

\item [] 3) The probabilities of: the initial proliferation of cancer
cells $p_0$; the effect of therapy on normal cells and cancer cells
($p_{drugn}$ and $p_{drugc}$).

\end{itemize}

The output data are the time series of the density of each type of
cell and the final configuration of the lattice at any time step. In
addition, the time-spatial configurations, controlled by a package
denominated g2 \cite{g2_manual} whose commands are inserted into the
computer program in FORTRAN, are generated in "real time". This
package may be used in C, PYTHON and PERL.

The following is a description of each subroutine based on the local
rules:

\begin{itemize}

\item [] a) {\bf Normal Cell (NoC)} - a random number $y$ is  compared
to $p_{drugn}$. If $y < p_{drugn}$, the growth potential $P_{noc}$
is confirmed: if $P_{noc} = 0$, the cell dies and the site becomes
empty; otherwise, it remains occupied by a normal cell but $P_{noc}
= 0$. If $y \ge p_{drugn}$ and if there is at least one neighboring
cancer cell, then the normal cell is 'dislocated', $P_{noc} = 0$ and
the site becomes empty; otherwise it remains a normal cell.

\vspace{0.3cm}

\item [] b) {\bf Cancer Cell (CC)} - if all of its neighbors are cancer cells and
its growth potential reaches a fraction $f$ of lattice size $L$, the
cancer cell becomes necrotic. Otherwise, a number $y$ is randomly
chosen. If $y < p_{drugc}$ and $t > t_{ap}$, the potential $P_{cc}$
is confirmed: if $P_{cc} > 0$, it is reduced by a unit and the cell
remains a cancer cell; otherwise the cell dies and the site becomes
empty. Finally, if $y \ge p_{drugc}$, the cell remains a cancer
cell; however, its growth potential $P_{cc}$ increases by a unit.

\vspace{0.3cm}

\item [] c) {\bf Necrotic Cell (NeC)} - if at least one of its neighbors is neither
a cancer cell nor a necrotic cell, it is eliminated and the site
becomes empty; otherwise, it continues necrotic.

\vspace{0.3cm}

\item [] d) {\bf Empty Site (ES)} - if there are cancer and normal cells in its
neighborhood, the local battle between cancer cells and normal cells
is such that if the sum of the potential growth of its neighboring
cancer cells is greater or equal to the sum of the potential growth
of its normal cell neighbors, then the empty site is occupied by a
cancer cell that diffuses from one of the randomly chosen neighbors;
otherwise, it is occupied by one of the randomly chosen normal cells
that were previously dislocated. If there are only normal cells in
its neighborhood, it becomes a normal cell through a process of
regeneration.  Finally, if none of its neighbors are cancer cells or
normal cells, it remains empty.

\end{itemize}

\section{RESULTS: SIMULATED TIME SERIES AND TIME-SPATIAL PATTERNS}
\label{sec3}

Computational simulations of the model were performed in order to
analyze two classes of behavior: cases in which no treatment was
given and treatment cases. In each subsection, the time series of
the simulated tumor as well as time-spatial patterns are shown.

\vspace{1cm}

\subsection{NO TREATMENT CASE}

\label{subsec31}

\begin{figure}
\begin{center}
\includegraphics*[width=6.0cm]{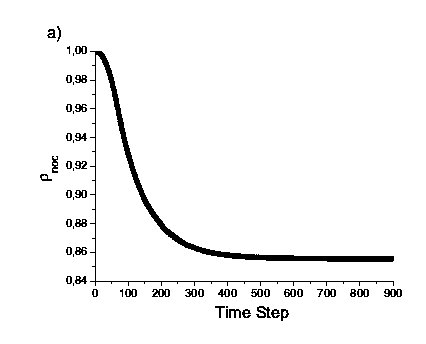}
\includegraphics*[width=6.0cm]{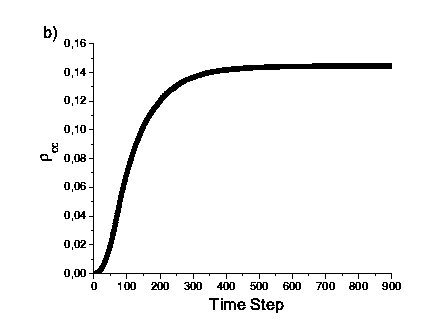}
\end{center}
\caption{\label{Figure1} Nonnecrotic tumor: the average of simulated
time series of the density of: a) normal cells, b) cancer cells. We
consider $M_{samples} = 200$ and the following parameter values: $L
= 251$, $p_{0} = 0.95$, $f = 0.6$, $p_{drugn} = p_{drugc} =
 0.0$.}
\end{figure}

In the case of no treatment, the following parameters are
considered: $p_{drugn} = p_{drugc} = t_{ap} = 0$. For fixed values
of $L$, $t_{final}$ and $p_0$, but different values of $f$, in
Figures \ref{Figure1} and \ref{Figure3}, the time series of the
density of cells for nonnecrotic and necrotic tumors, respectively,
are shown. In both cases, the cell densities reach saturated values
due to the effects of the competition between normal cells and
cancer cells, and the time-dependent mitotic probability. Comparing
Figures \ref{Figure1}b and \ref{Figure3}b, the stationary value of
cancer cell density is clearly greater in necrotic tumors than in
nonnecrotic ones. This is a consequence of the fact that $\Delta
p_{mitot}(t)$ assumes smaller values in necrotic tumors compared to
nonnecrotic ones because it does not depend on the density of
necrotic cells. In both cases, the average of the different samples
is considered, corresponding to different seeds of random numbers.

\begin{figure}
\begin{center}
\includegraphics[width=6.0cm]{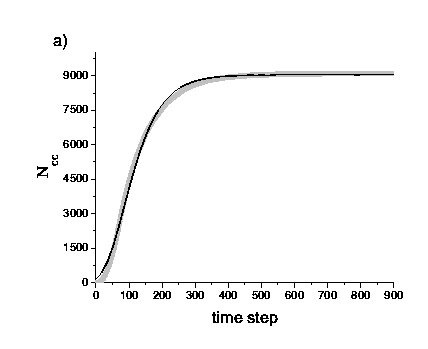}
\includegraphics[width=6.0cm]{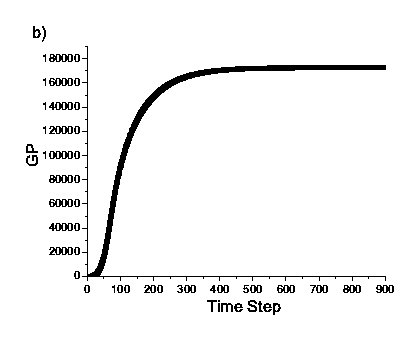}
\end{center}
\caption{\label{Figure2} Non-necrotic case: assuming
$M_{samples}=200$ and the same parameters values of Figure
\ref{Figure1}, the average of simulated time series of: a) the
number of cancer cells (grey color); b) the growth potential of
cancer cells (black color). The Gompertzian fitting is applied to
(a) (black color) those in which parameters of time series are
$\alpha_0=(6.58 \pm 0.09) \times 10^{-2}$, $\beta=(1.631 \pm 0.008)
\times 10^{-2}$ and $n_0=(1.6 \pm 0.06) \times 10^{2}$.}
\end{figure}

\begin{figure}
\begin{center}
\includegraphics*[width=6.0cm]{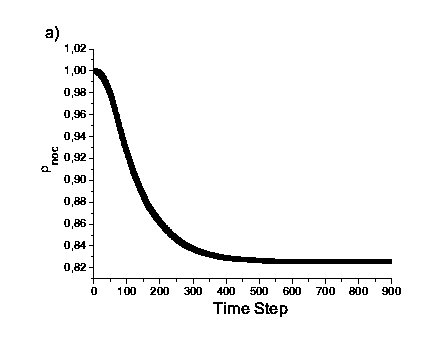}
\includegraphics*[width=6.0cm]{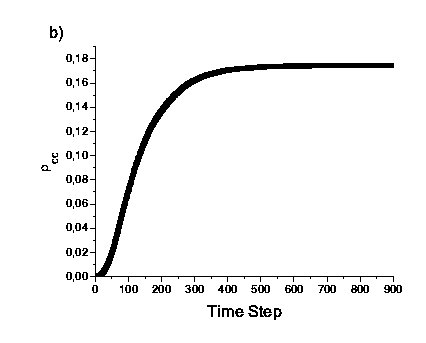}
\end{center}
\caption{\label{Figure3} Necrotic case: the average of simulated
time series of density of  a) normal cells, b) cancer cells. We
consider $M_{samples}=200$ and the following parameter values:
$L=251$, $p_0=0.95$, $f=0.2$, $p_{drugn}=0.0$, $p_{drugc}=0.0$.}
\end{figure}

\begin{figure}
\begin{center}
\includegraphics*[width=6.0cm]{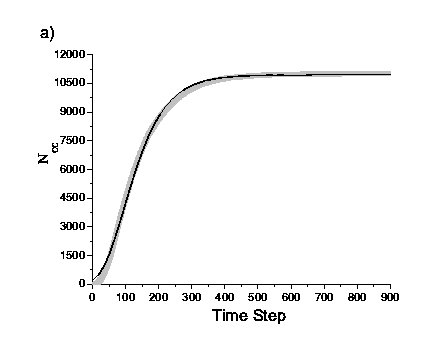}
\includegraphics*[width=6.0cm]{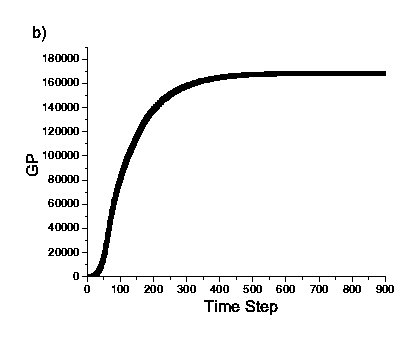}
\end{center}
\caption{\label{Figure4} Assuming $M_{samples}=200$ and the same
parameters values of Figure \ref{Figure3}, the average of simulated
time series of: a) the number of cancer cells (grey color); b) the
growth potential of cancer cells (black color). The Gompertzian
fitting is applied on (a) (black color) those in which parameters of
time series are $\alpha_0=(5.86 \pm 0.06) \times 10^{-2}$,
$\beta=(1.450 \pm 0.006) \times 10^{-2}$ and $n_0=(1.926 \pm 0.005)
\times 10^{2}$.}
\end{figure}

Figures \ref{Figure2}a and \ref{Figure4}a  show that the tumor
growth obeys the Gompertzian function both in nonnecrotic and
necrotic cases. This behavior is observed with respect to the number
of cancer cells for a range of values of $p_0$. In Gompertzian
growth, the specific growth rate of the number of cancer cells
decreases logarithmically:

\begin{equation}
\label{gompertzrate}
 \frac{1}{n_{cc}}\frac{dn_{cc}}{dt} = \alpha_0
- \beta \log \left(\frac{n_{cc}}{n_0}\right)
\end{equation}

\noindent where

 \begin{itemize}
 \item [] a) $n_0$ is the initial population of cancer cells;
 \item [] b) $\alpha_0$ is the specific growth rate of $n_0$ cells
 at $t=0$;
 \item [] c) $\beta$ measures how rapidly the curve departs from a
 singular exponential and curves over, assuming its characteristic
 shape.
\end{itemize}

 The solution of (\ref{gompertzrate}) is

 \begin{equation}
 \label{gompertzsolution}
 n_{cc}(t)=n_0 \exp \left\{\frac{\alpha_0}{\beta}[1-\exp(-\beta t)]\right\}.
 \end{equation}

It is evident that the stationary value of $n_{cc}$ is
$n_{{cc}_{\infty}}=n_0 \exp(\alpha_0/\beta)$. According to the
Gompertzian fitting represented by equation \ref{gompertzsolution},
the results of the simulations shown in Figures
 \ref{Figure2} and \ref{Figure4} correspond respectively to the following parameters:

\begin{itemize}
\item [] I) The nonnecrotic tumor: $\alpha_0=(6.58 \pm 0.09) \times 10^{-2}$, $\beta=(1.631 \pm
0.008) \times 10^{-2}$ and $n_0=(1.60 \pm 0.06) \times 10^{2}$

 \item [] II) The necrotic tumor:  $\alpha_0=(5.86 \pm 0.06) \times
10^{-2}$, $\beta=(1.450 \pm 0.006) \times 10^{-2}$ and $n_0=(1.926
\pm 0.005) \times 10^{2}$
\end{itemize}

Comparing the above parameters of (I) and (II), the behavior of
nonnecrotic and necrotic tumors was found to be very similar. It was
also found that the number of cancer cells obeys the Gompertzian
fitting.

An important confirmation of our model is shown by comparing the
Gompertzian fitting parameters of simulated tumors with the
corresponding parameters of actual tumors \cite{Demicheli}
\cite{Brunton}. For instance, in the case of the testicular tumors
shown in Table 3 of reference \cite{Demicheli}, the $\beta$ values
are in the range of $[0.005; 0.016]$ $day^{-1}$. Our simulated
$\beta$ values are within the above range if we consider the time
step of our simulations to be one day.

In relation to the parameters $\alpha_{0}$ and $n_0$, the simulated
values are not comparable to the actual values, since no information
on the initial size of the tumor was included in our model. Both
$\alpha_0$ and $n_0$ are strongly dependent on that information.

\begin{figure}
\begin{center}
\includegraphics[width=0.4\textwidth,angle=0]{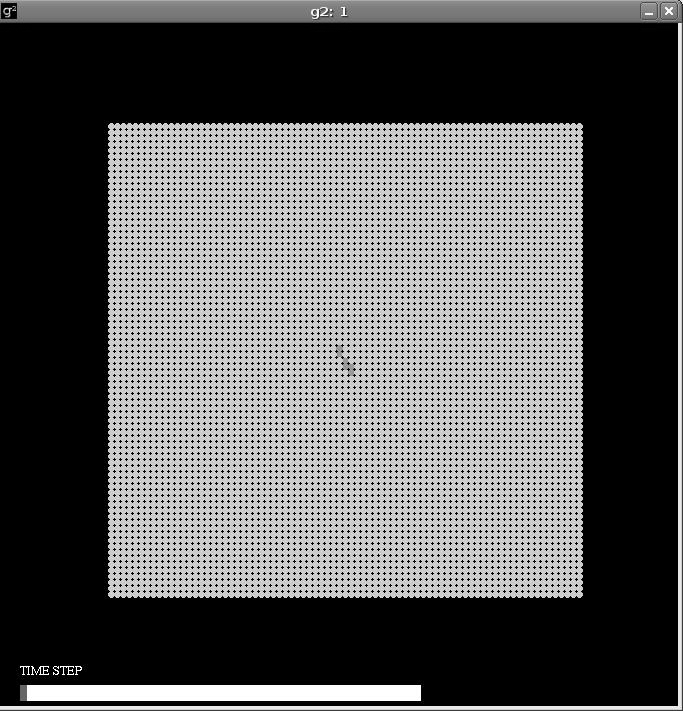}
\includegraphics[width=0.4\textwidth,angle=0]{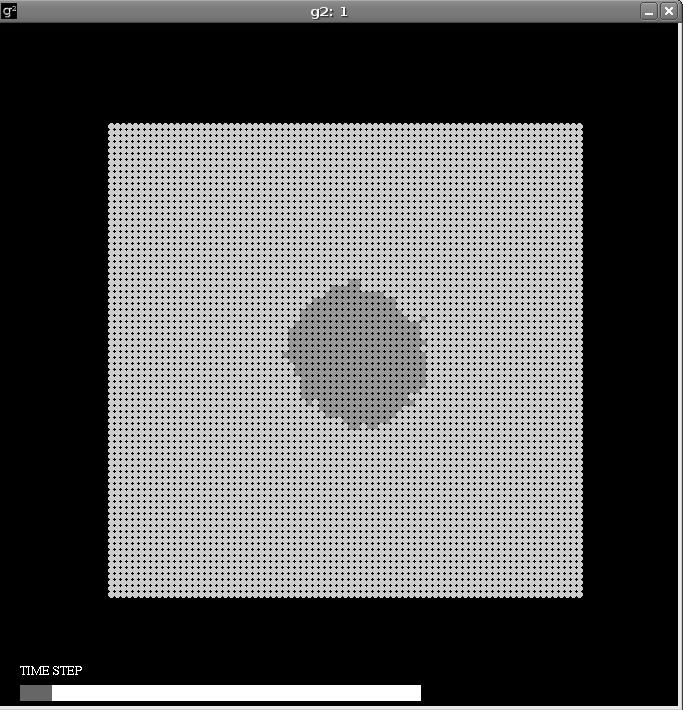}
\vspace{1cm}
\includegraphics[width=0.4\textwidth,angle=0]{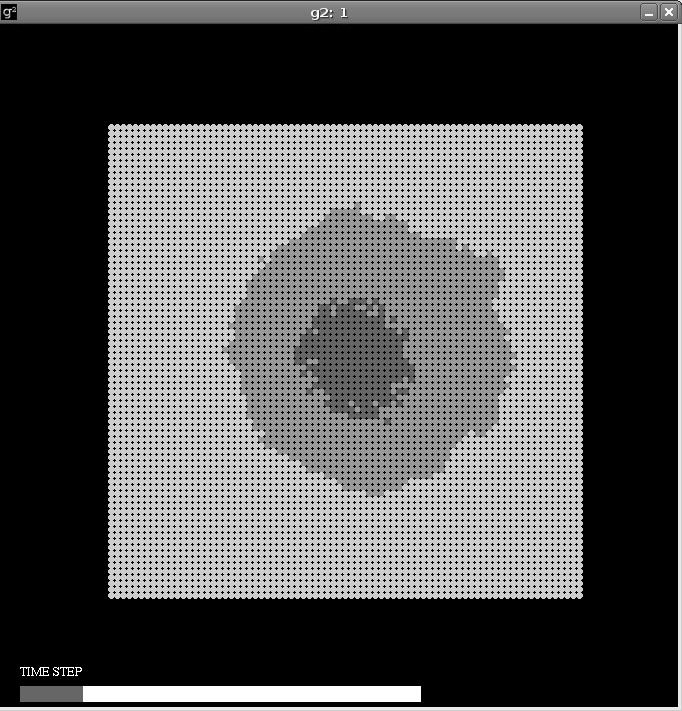}
\includegraphics[width=0.4\textwidth,angle=0]{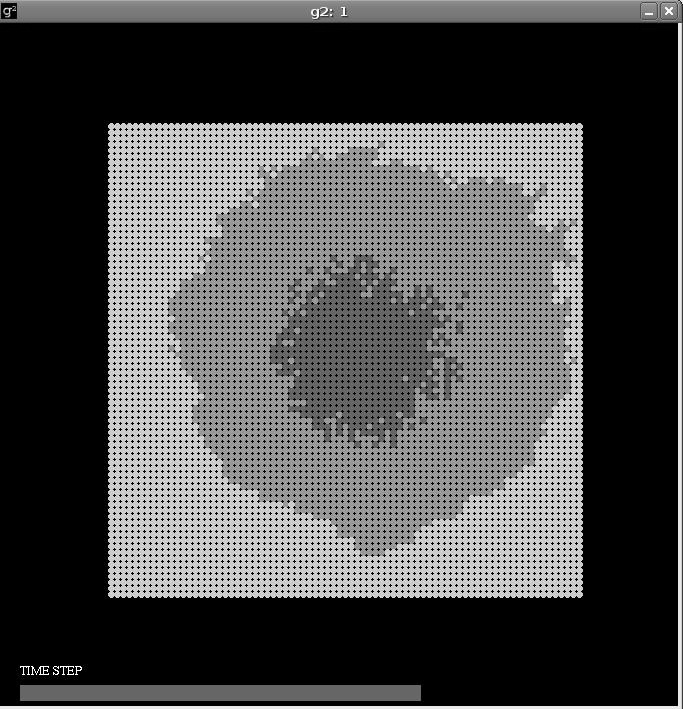}
\end{center}
\caption{\label{Figure5} The spatial distribution of states at
consequent time steps (see the time step bar) using the same
parameters values of Figure \ref{Figure3} except $L=81$ and
$p_0=0.8$. The final time step is $t_{final}=600$. Dark colors
(light grey, grey, and dark grey) correspond, respectively,
to normal, cancer, and necrotic cells.}
\end{figure}

Finally, it would be very interesting to discover whether it is
possible to relate the cellular automata no-therapy parameters $L$,
$f$ and $p_0$ with the Gompertzian parameters for the range of
values of $p_0$ and to assess whether the number of cancer cells
obeys a Gompertzian growth pattern. In this case, some preliminary
conclusions may be drawn with respect to $n_\infty$: it decreases in
accordance with the necrotic parameter $f$ but it increases linearly
as a function of lattice size $L$ and, exponentially with $p_0$.
Concerning the other Gompertzian parameters, the answer to this
question is not so simple. Following these conclusions leads us to a
much more interesting quandary if we want to make the model more
realistic: to estimate realistic ranges of the CA no-therapy
parameters $L$, $f$ and $p_0$ based on the Gompertzian parameters of
actual solid tumors.

With respect to the time-spatial patterns of the simulated tumors,
Figure \ref{Figure5} shows the lattice configuration at some time
during the steady state in the necrotic (Figure \ref{Figure5}b)
cases. Using the g2 package, it was found that, although tumor
growth leads to the compact shape that is characteristic of solid
tumors, the growth process is such that its boundary is irregular
for any time $t$, as, for example, for the time steps represented in
Figure \ref{Figure5}. A similar type of behavior is observed for the
time-spatial patterns in the nonnecrotic tumor. In relation to the
process of necrosis, the necrotic region was found to be inside the
tumor at any time $t$ \cite{Kansal} \cite{Bellomo}, as shown in
Figure \ref{Figure5}.

\subsection{TREATMENT CASE}
\label{subsec32}

\begin{figure}
\begin{center}
\includegraphics[width=6.0cm]{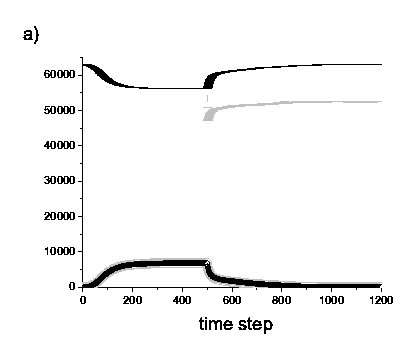}
\includegraphics[width=6.0cm]{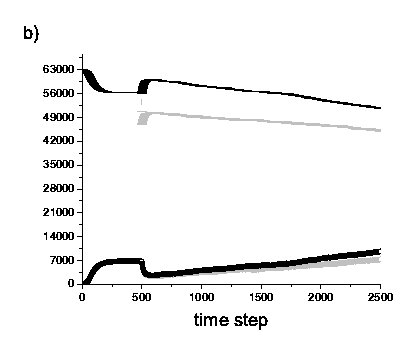}
\end{center}
\caption{\label{Figure6}a) Successful treatment: the average of
simulated time series of the number of normal cells (-) and cancer cells (circle) assuming
$M_{samples}=200$ and the following parameters values: $L=251$,
$p_0=0.8$, $f=0.2$, $p_{drugc}= 0.9$, $t_{ap}= 500$, and two values
for $p_{drugn}$: $10^{-4}$ (dark grey), and $10^{-2}$ (light grey);
b) Unsuccessful treatment: the time series of the number of normal
cells (-) and cancer cells (circle) assuming $M_{samples}=200$ and the following
parameters values: $L=251$, $p_0=0.8$, $f=0.2$, $p_{drugc}= 0.7$,
$t_{ap}= 500$, and two values for $p_{drugn}$: $10^{-4}$ (dark
grey), and $10^{-2}$ (light grey).}
\end{figure}

Figures \ref{Figure6}a and \ref{Figure6}b show the simulated time
series of the number of cancer cells and normal cells for the
different values of $p_{drugc} > p_0$ and $p_{drugc} < p_0$
corresponding to successful treatment (cure) and unsuccessful
treatment (non-cure), respectively. In each figure, two values of
$p_{drugn}$ are considered. It can be clearly seen that, for both
values of $p_{drugn}$, the success of the treatment does not change.
It is also clear that the tumor size increases until tap (see
figures \ref{Figure6}a and \ref{Figure6}b). However when $p_{drugc}
> p_0$, the reduction in tumor size starts, as expected, after $t_{ap}$ time steps.

When treatment is successful ($p_{drugc} > p_0$), it was found that,
for an upper value of $p_{drugn}$ ($p_{drugn} = 10^{-2}$; light
grey), more normal cells are eliminated compared with a lower value
of $p_{drugn}$ - $p_{drugn} = 10^{-4}$ as shown in Figure
\ref{Figure6}a (dark grey). Therefore, we may conclude that very
small values of $p_{drugn}$ correspond to optimal therapy. However,
in the case of unsuccessful treatment ($p_{drugc} < p_0$), when
$p_{drugn}$ is increased, both a decrease in the number of normal
cells and a slower rate of increase of cancer cells is found (see
fig \ref{Figure6}b).

A systematic analysis of the parameter space presented in the next
section will provide a more precise conclusion about the role of
$p_{drugn}$ parameter.

\begin{figure}
\begin{center}
\includegraphics[width=0.4\textwidth,angle=0]{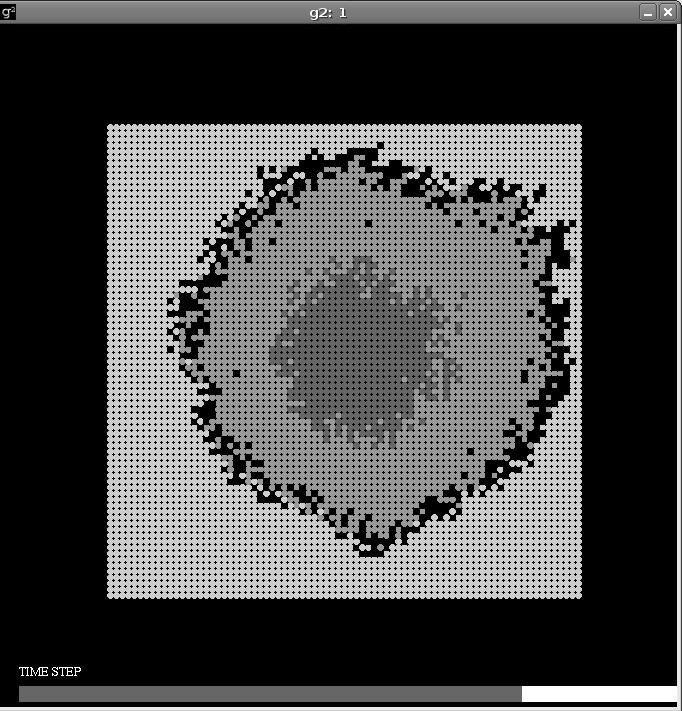}
\includegraphics[width=0.4\textwidth,angle=0]{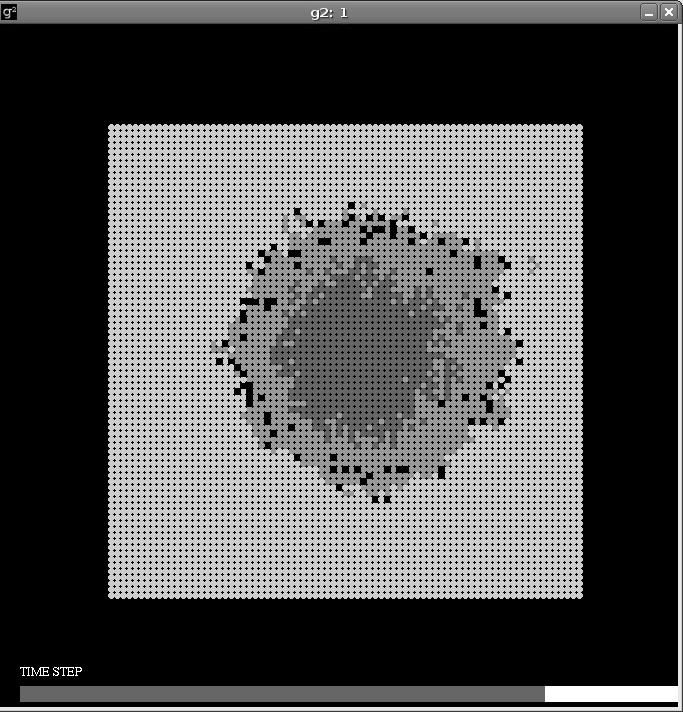}
\vspace{1cm}
\includegraphics[width=0.4\textwidth,angle=0]{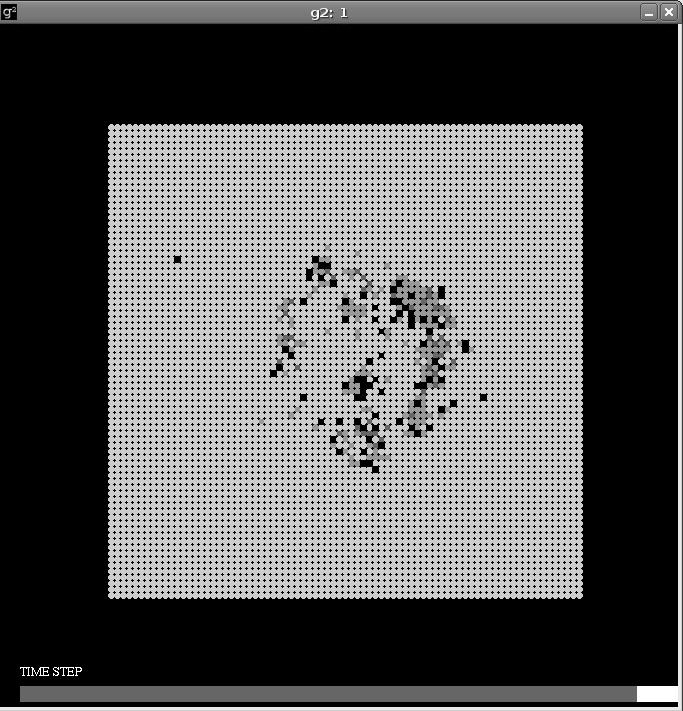}
\includegraphics[width=0.4\textwidth,angle=0]{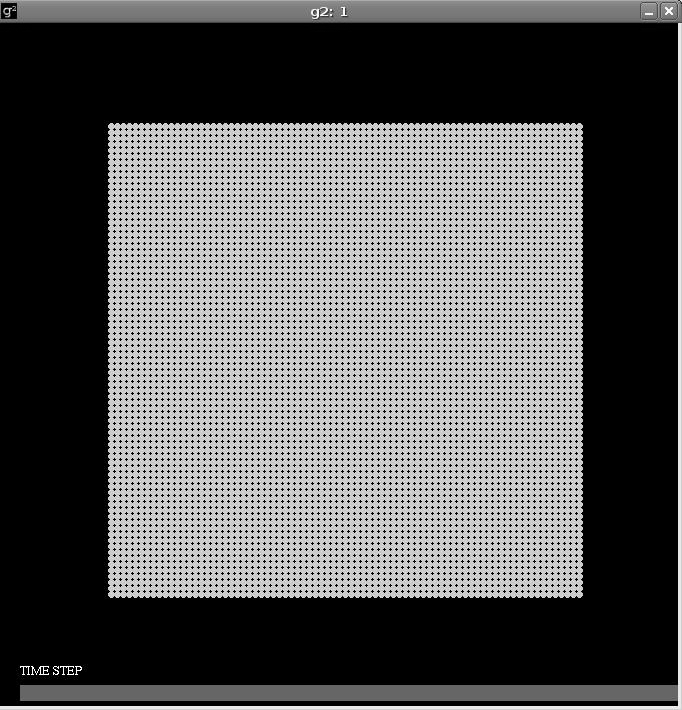}
\vspace{1cm}
\end{center}
\caption{\label{Figure7} The spatial distribution of states at
consequent time steps (see the time step bar) using the same
parameters values of Figure \ref{Figure6}a except $L=81$. The final
time step is $t_{final}=1400$. Dark colors (light grey, grey, dark
grey, and black) correspond, respectively, to normal, cancer,
necrotic cells, and empty sites.}
\end{figure}

In order to analyze the effect of therapy on cancer cells, the
time-spatial distribution of the simulated tumors was followed using
the g2 package. Figure \ref{Figure7} shows the configuration of the
lattice at different time steps from the beginning of therapy until
the tumor is eliminated.

\section{RESULTS: PARAMETER SPACE}
\label{sec4}

Analysis of the parameter space is important in order to confirm the
robustness of the model and to identify the most relevant parameters
for the dynamics of the model.

According to the relevance of some CA parameters to the features of
the phenomenon, this analysis of parameter space was divided into
two parts: the occurrence of necrosis (no treatment cases) and
reaching a state of cure (treatment cases).

\subsection{THE OCCURRENCE OF NECROSIS}
\label{subsec41}

\begin{figure}[!h]
\begin{center}
\includegraphics[width=6.0cm]{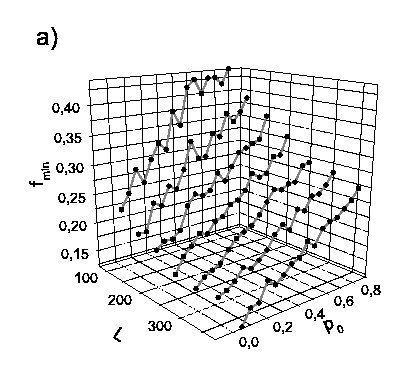}
\includegraphics[width=6.0cm]{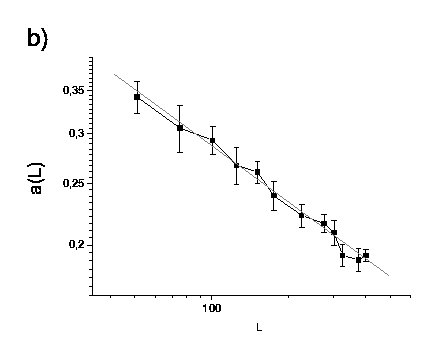}
\end{center}
\caption{\label{Figure8} a) Parameter space in the case of no
treatment: $f^{min} \times p_0 \times L$; the following parameters
values are fixed: $p_{drugn}=0$, $p_{drugc}=0$; $t_{final}=5000$. b) the dependence of
the angular coefficient of $f_{min}$, $a(L)$, with $L$.}
\end{figure}

In the first part, the parameters are again established as:
$p_{drugn} =  p_{drugc} = t_{ap} = 0$ with the aim of evaluating the
effect of necrosis, and the minimum value of f is obtained for each
pair of values $(L; p_0)$. The values of $L$ are presumed to range
from 101 to 251, increasing the interval by $\Delta L = 25$. With
respect to $p_0$, the whole interval from 0.1 to 0.9 is taken into
account, increasing $\Delta p_{mitot}(t) = 0.1$.  Figure
\ref{Figure8} shows that a transition exists between the nonnecrotic
(lower) and necrotic (upper) regions of parameter space.

It was found that:

\begin{equation}
\label{fmin} f_{min}=a(L) p_0 + b(L)
\end{equation}

where the linear and the angular coefficients, $b(L)$ and $a(L)$ are
different for distinct values of the lattice size.

The angular coefficient $a(L)= a_0 L^\gamma$ where $a_0=1.12$ and
$\gamma=0.29$ (see figure \ref{Figure8}b).

Since the lattice size is an intrinsic parameter of the model, that
transition is such that the the maximal growth probability $p_0$ is
the control parameter, while the necrotic parameter $f$ is the order
parameter.

\subsection{REACHING THE CURE STATE}
\label{subsec42}

If the tumor is submitted to systemic treatment represented by a
probability $p_{drugc} \ne 0$ starting at $t_{ap} \ne 0$ that
affects normal cells with a probability $p_{drugn} < p_{drugc}$, our
aim is to establish the minimum value of $p_{drugc}$ in order to
achieve a state of cure, i.e., no cancer cells.

The analysis is now more complex than the one performed in
subsection \ref{subsec41} in the sense that there are 4 important
parameters that control the behavior of the therapeutic effect:
$p_{drugc}$, $p_{drugn}$, $t_{ap}$, and $p_{0}$. The relevance of
$p_0$ is evident in subsection \ref{subsec32}. This analysis is
divided into two parts.

\begin{figure}
\begin{center}
\includegraphics[width=6.0cm]{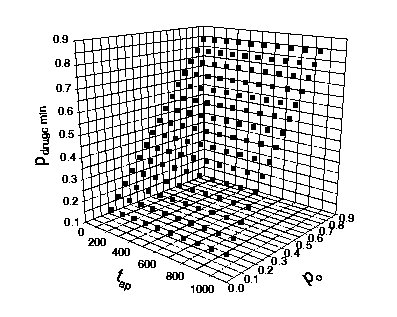}
\end{center}
\caption{\label{Figure9} Parameter subspace of treatment case:
$p_{drugc}^{min} \times p_0 \times t_{ap}$; the following parameters
values are fixed: $L=251$, $f=0.6$, $p_{drugn}=10^{-4}$;
$t_{final}=5000$.}
\end{figure}

In the first part, motivated by the behavior observed in figure
\ref{Figure6}a, the parameter is defined as $p_{drugn}$, assuming
very small values ($p_{drugn} = 10^{-4})$ to simulate optimal
therapy. Lattice size is also established as $L = 251$ and the
necrotic parameter $f = 0.6$. Analogously to the method applied in
subsection \ref{subsec41}, the minimum value of $p_{drugc}$ was obtained
for each pair of values $(p_0; t_{ap})$. The range of values for
$p_0$ and tap are, respectively, $[0.1; 0.9]$ and $[0; t_{final}]$.
Figure \ref{Figure9} shows the relevant role of $p_0$ dividing the
parameter subspace into two regions: the lower is the non-cure state
while the upper corresponds to the state of cure. In this case $p_0$
is again the control parameter but $p_{drugc}$ is the order
parameter.

\begin{figure}
\begin{center}
\includegraphics[width=6.0cm]{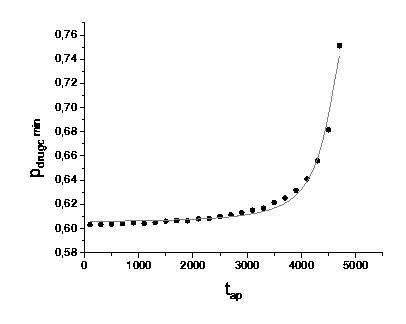}
\end{center}
\caption{\label{Figure10} Parameter subspace of treatment case:
$p_{drugc}^{min}  \times t_{ap}$; the following parameters values are
fixed: $L=251$, $f=0.6$, $p_0=0.6$; $p_{drugn}=10^{-4}$;
$\delta_t=t_{final}-t_{ap}=500$. The Lorentzian fitting (grey color) parameters
are $A_0=0.60$, $A_1=186.61$, $A_2=4829.92$, $A_3=776.29$.}
\end{figure}

In the second part, a value of $p_0 = 0.8$ remains fixed, while the
pair varies $(p_{drugn}; t_{ap})$, again in order to obtain the
minimum value of $p_{drugc}$ that would be sufficient to effectively
eliminate the tumor, now related to the effect of the drug on the
normal cells. The whole interval of $p_{drugn}$ was taken into
account from very small values $10^{-4}$ until $0.9$. It would be
necessary to extend $t_{final}$ in order to maintain the duration of
application, since in this analysis the parameter $t_{ap}$ varied.
Since this behavior is similar for any value of $p_{drugn}$, in
figure \ref{Figure10} the minimum value of $p_{drugc}$ is shown to
increase as a function of $t_{ap}$ in accordance with a Lorentzian
function:
$$p_{drugc}^{min} = A_0 + \frac{2A_1}{\pi}\,\frac{A_3}{4(t_{ap}-A_2)^2 + A_3^2}$$

This result means that the minimal rate of infusion of the drug to
eliminate the tumor has to be greater if the treatment begins later.
It emphasizes how important it is to initiate treatment as early as
possible in order to reduce the infusion rate of the drug.

\begin{figure}
\begin{center}
\includegraphics[width=6.0cm]{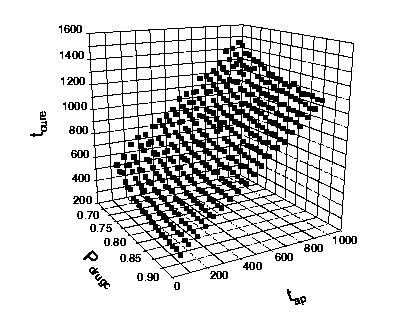}
\end{center}
\caption{\label{Figure11} Analysis of cure time in the treatment
case: $t_{cure} \times p_{drugc}^{min} \times t_{ap}$; the following parameters values
are fixed: $L=251$, $f=0.6$, $p_0=0.6$, $p_{drugn}=10^{-4}$, $t_{final}=500$.}
\end{figure}

Finally in the cases in which the tumor is eliminated ($p_{drugc} >
p_0$), for fixed values of $L$, $f$, $p_0$ and $p_{drugn}$, the cure
time ($t_{cure}$) is estimated in relation to the initial time
application $t_{ap}$, and $p_{drugc}$ (see Figure \ref{Figure11}).
Note that $t_{cure}$ is not a parameter but a consequence of the
time evolution of the system. Figure ref{Figure11} shows that
$t_{cure}$ increases linearly with $t_{ap}$ with an angular
coefficient equal to 1. Figure \ref{Figure11} also shows that, for a
fixed $t_{ap}$, $t_{cure}$ decreases with $p_{drugc}$. This result
also shows how important it is to start treatment as early as
possible to reduce the amount of time required to reach the state of
cure.

\section{DISCUSSION AND CONCLUDING REMARKS}
 \label{sec5}

The model proposed in this study is capable of capturing the
Gompertzian behavior of avascular tumor growth. The competition
between normal and cancer cells and the dynamic character of the
mitotic probability are the relevant components of the success of
this model.

The number of cancer cells simulates tumors {\it in vitro} due to
the two-dimensional character of the model, and their potential
growth simulates a tumor in vivo due to its three-dimensional
nature. Figures \ref{Figure2} and \ref{Figure4} show that the model
is able to capture the dynamics of both {\it in vivo} and {\it in
vitro} avascular tumor growth. The simulated values for the most
important Gompertzian parameter $\beta$, which characterizes the
Gompertzian shape, are compatible with the parameter values of some
tumors \cite{Demicheli} \cite{Brunton}.

The model is also able to capture necrotic and nonnecrotic tumors
depending on the values of the parameter $f$.  It is well-known that
necrosis, unlike apoptosis, is a typical phenomenon found in a group
of cells that is simulated in our model by the changing of the
potential growth of the cells.

The time-spatial patterns reveal a tumor with a compact shape and
irregular boundaries, as occurs in some solid tumors \cite{Patel},
\cite{Ferreira}. Evidence was also found to confirm the three stages
of avascular tumor growth \cite{Adam}:

\begin{itemize}

 \item [] a) Stage I - when the tumor grows exponentially due to available
 resources (nutrients and oxygen) - see first stage in Figure \ref{Figure5};

 \item [] b) Stage II - when the stabilization of $\Delta p_{mitot}$ starts but there are
  still enough resources to ensure that necrosis does not occur - see second stage in Figure \ref{Figure5}b;

 \item []  c) Stage III - when necrosis
   may occur depending on the value of the parameter $f$ because the resources are
   insufficient to provide for tumor growth - see third stage in Figure \ref{Figure5}c in the case of necrosis.

   \end{itemize}

The next stage, the angiogenic phase of the tumor \cite{Folkman},
which has not yet been dealt with in our model, corresponds to
vascular growth. To be able to evaluate this stage, the process of
angiogenesis would have to be taken into account \cite {Adam}.

In the case of no treatment ($p_{drugc} = p_{drugn} = 0$), the
minimum value of $f$ for necrosis to occur is governed by the
equation (\ref{fmin}) that was obtained from the investigation of
the parameter space. This means that, for a simulated tissue of
dimension $L$, the occurrence of necrosis depends linearly on the
maximum value of mitotic probability of the specific tumor.

In relation to the case in which therapy was implemented, a
continuous strategy of systemic therapy, i.e. chemotherapy, was
selected. Although the schedule of chemotherapy is usually periodic,
in this case it was decided to simulate this less realistic
situation so that the effect of the parameters $p_0$ and $p_{drugc}$
could be compared in a simple fashion. It was thus found that, with
respect to avascular tumor growth, when the values of the parameter
$p_{drugn}$ are very small and when $p_{drugc}$ is slightly greater
than $p_0$, the tumor is completely eliminated (see Figure
\ref{Figure7}). Since neovascularization has not yet been
triggered, a state of cure may be expected to occur with a periodic
schedule.

The time-spatial patterns of the cases in which therapy was
implemented (see Figure \ref{Figure8}) show that the drug acts from
the borders of the simulated solid tumor inwards, as would be
expected in the case of solid tumors.

With respect to the effect of parameters on the state of cure, for
fixed values of lattice size $L$ and necrotic parameter $f$, and for
a range of values of $t_{ap}$ and $p_{drugn}$, the minimum value of
$p_{drugc}$ again coincides with $p_0$ (see Figure \ref{Figure9}).
This reinforces the importance of the higher value of $p_0$ in the
response of the tumor to therapy. It retains the memory of cancer
cells with respect to the onset of mitosis.

Finally, Figures \ref{Figure10} and \ref{Figure11} illustrate a very
relevant finding from the phenomenological point of view: the
importance of initiating therapy as early as possible in order to
reach the state of cure. The cure time was found to be proportional
to $t_{ap}$ that measures the instant when the infusion starts (see
Figure \ref{Figure11}) and the minimum value of the therapeutic
infusion to eliminate the tumor increases nonlinearly as a function
of the starting point of therapy.

Future studies should be carried out to generalize the model with
the objective of including the angiogenic process and the periodic
schedule of systemic therapy. However, the most important
perspective of this line of investigation is to compare the model
with the in vitro tumor growth of cells from specific tissue samples
and to compare parameter values. It would then be possible to relate
the Gompertzian fitting parameters with the parameters of the model.

\vspace{1cm}

{\bf Acknowledgements:} The authors would like to thank Ramon
El-Bach\'a for his very useful discussions on the process of tumor
growth and Nelson Alves Jr. for his valuable collaboration at the
beginning of this study and his help in manipulating the time
spatial patterns. This work is partially supported by CNPq --
Conselho Nacional de Desenvolvimento Cient\'{\i}fico e Tecnol\'ogico
(Brazilian Agency).


\begin{thebibliography}{99}

\bibitem{Who} The World Health Report 2006: Working Together for Health (World Health
Report).

\bibitem{Preszosi} L. Preziosi (editor), {\it Cancer Modelling and Simulation},
 Chapman \& Hall/CRC, London, 2003.

\bibitem{MMC}  E. Y. Rodin (editor-in-chief), A. Y. Yakovlev and S. H. Moolgavkar
(guest editors), {\it Modeling and data analysis in cancer studies},
Math. Comp. Mod. {\bf 33}, n. 12-13, 2001.

\bibitem{Wheldon} T. E. Wheldon, {\it Mathematical Models in
Cancer Research}, Adam Hilger, Bristol, 1988.

\bibitem{Laird} A. K. Laird, Dynamics of tumour growth: comparison of growth rates and extrapo
lation of growth curves to one cell. Br. J. Cancer {\bf 19} (1965)
278-291.

\bibitem{Demicheli} R. Demicheli, Growth of Testicular Neoplasm Lung Metastases:
Tumor-Specific Relation between two Gompertzian Parameters. Eur. J.
Cancer {\bf 16} (1980), 1603-1608

\bibitem{Brunton} G. F. Brunton and T. E. Wheldon, The Gompertz equation and the
construction of tumor growth curves. Cell Tissue Kinet. {\bf 13}
(1980), 455-460.

\bibitem{Clare} S. E. Clare, F. Nakhlis, and J. C. Panetta, Molecular biology of breast
cancer metastasis: The use of mathematical models to determine
relapse and to predict response to chemotherapy in breast cancer.
Breast Cancer Res. {\bf 2} (2000), 430�435.

\bibitem{Spratt} J. A. Spratt, D. von Fournier, J. S. Spratt, E. E.
Weber, Decelarating Growth and Human Breast Cancer. Cancer {\bf 71}
(1993), 2013-2019.

\bibitem{Bellomo} N. Bellomo and L. Preziosi, Modelling and Mathematical Problems
Related to Tumor Evolution and Its Interaction with the Immune
System. Mathl. and Comput. Modelling {\bf 32} (2000), 413-452.

\bibitem{Guiot} C. Guiot, P. G. Degiorgis, P. P. Delsanto, P. Gabriele, T. S.
Deisboeck, Does tumor growth follow a "universal law"?. J. Theor.
Biol. {\bf 225} (2003), 147-151.

\bibitem{Patel} A. A. Patel, E. T. Gawlinskia, S. K. Lemieuxe, and
R. A. Gatenbyb, A Cellular Automaton Model of Early Tumor Growth and
Invasion*: The Effects of Native Tissue Vascularity and Increased
Anaerobic Tumor Metabolism. J. theor. Biol. {\bf 213} (2001),
315-331.

\bibitem{Castro} M. A. A. Castro, F. Klamt, V. A. Grieneisen, I.
Grivicich, and J. C. F. Moreira, umour cell phenotype Gompertzian
growth pattern correlated with phenotypic organization of colon
carcinoma, malignant glioma and non-small cell lung carcinoma cell
lines. Cell Prolif. {\bf 36} (2003) 65�73.

\bibitem{Galam} S. Galam, and J. P. Radonski, Cancerous tumor:
the high frequency of a rare event, Phys. Rev. E {\bf 63} (2001)
051907.

\bibitem{Shen} AN-S. Qi, X. Zheng, C-Y. Du, and B-S. An, A cellular
automaton model of cancerous growth, J. Theor. Biol. {\bf 161}
(1993) 1-12.

\bibitem{Dormann} S. Dormann and A. Deutsch, Modeling of self-organized
avascular tumor growth with a hybrid cellular automaton. In Silico
Biology {\bf 2} (2002), 0035.

\bibitem{Gerlee} P. Gerlee and A.R.A. Anderson, An evolutionary hybrid cellular automaton model
of solid tumour growth, J. Theor. Biol. {\bf 246} (2007) 583�603.

\bibitem{Rygaard} K. Rygaard and M Spang-Thomsen, Comparative Gompertzian Analysis of Alterations
of Tumor Growth Patterns, Cancer Res. {\bf 54} (1994) 4385-4392.

\bibitem{Martin} R. Martin and K. L. Teo, {\it Optimal control of
drug administration in cancer chemotherapy}, World Scientific,
Singapore, 1994.

\bibitem{Swan} G. W. Swan, Role of Optimal Control Theory in Cancer
Chemotherapy, Math. Biosci. {\bf 101} (1990), 237-284.

\bibitem{Murray} J. M. Murray, Optimal Drug Regimens in Cancer
Chemotherapy for Single Drugs that Block Progression through the
Cell Cycle, Math. Biosci. {\bf 123} (1994), 183-193 (1994).

\bibitem{Costa} M. I. S. Costa, J. L. Boldrini and R. C.
Bassanezi, Chemotherapeutic Treatments Involving Drug Resistance and
Level of Normal Cells as a Criterion of Toxicity, Math. Biosci. {\bf
125} (1995), 211-228.

\bibitem{Matveev} A. S. Matveev and A. V. Savkin, Optimal
chemotherapy regimens: influence of tumors on normal cells and
several toxicity constraints, IMA J. Math. Appl. Med. Biol. {\bf 18}
(2001), 25-40.

\bibitem{Adam} J. A. Adam and S. A. Maggelakis, Diffusion regulated growth characteristics of
a spherical prevascular carcinoma, Bull. Math. Biol. {\bf 52} (1990)
549-582.

\bibitem{Wolfram} S. Wolfram, {\it Cellular Automata and Complexity}.
Addison-Wesley Publishing Company, New York, 1994.

\bibitem{Thecell} B. Alberts, A. Johnson, J. Lewis, M. Raff, K, Roberts, and P. Walter
 {\it Molecular biology of the cell}, Taylor and Francis Books, New York, 2002.

 \bibitem{Lobato} S.C. Ferreira Junior, M.L. Martins, and M.J.
Vilela, Morphology transitions induced by chemotherapy in carcinomas
in situ, Phys. Rev. E {\bf 67} (2003), 051914.

\bibitem{g2_manual} Milanovic Lj., Wagner, H. , g2 - graphic library (C) 1999 (http://g2.sourceforge.net)

\bibitem{Kansal} A. R. Kansal, S. Torquato, E. A. Chiocca, and T. S. Deisboeck, Emergence of a subpopulation
 in a computational model of tumor growth., J. Theor. Biol. {\bf 207} (2000), 431-441.

\bibitem{Ferreira} S.C. Ferreira Junior, M.L. Martins, and M.J.
Vilela, A growth model for primary cancer. Physica A {\bf 261}
(1998) 569-580.

\bibitem{Folkman} J. Folkman, Tumor Angiogenesis: Therapeutic implications.
N. Engl. J. Med. {\bf 285} (1971) 1182�1186.

\end{thebibliography}
\end{document}